\documentclass[twocolumn,nofootinbib,showpacs,prd,superscriptaddress]{revtex4}

\usepackage{graphics,graphicx}
\usepackage{amsmath}
\usepackage{psfrag}
\usepackage{dcolumn}
\usepackage{psfrag}

\usepackage{color}

\def\no{\nonumber \\}

\begin{document}

\title{
Comparison between numerical relativity and a new class of post-Newtonian 
gravitational-wave phase evolutions: the non-spinning equal-mass case
} 

\author{Achamveedu Gopakumar}
\affiliation{Theoretisch-Physikalisches Institut, Friedrich-Schiller-Universit\"at Jena,
Max-Wien-Platz 1,07743 Jena, Germany}
\author{Mark Hannam}
\affiliation{Theoretisch-Physikalisches Institut, Friedrich-Schiller-Universit\"at Jena,
Max-Wien-Platz 1,07743 Jena, Germany}
\author{Sascha Husa}
\affiliation{Max-Planck-Institut f\"ur Gravitationsphysik, Albert-Einstein-Institut, 
Am M\"uhlenberg 1, 14476 Potsdam,  Germany}
\author{Bernd Br\"ugmann}
\affiliation{Theoretisch-Physikalisches Institut, Friedrich-Schiller-Universit\"at Jena,
Max-Wien-Platz 1,07743 Jena, Germany}

\begin{abstract}
We compare the phase evolution of equal-mass nonspinning black-hole binaries from
numerical relativity (NR) simulations with post-Newtonian (PN) results obtained from 
three PN approximants: the TaylorT1 and T4 approximants, for which NR-PN comparisons
have already been performed in the literature, and the recently proposed
approximant TaylorEt.  
The accumulated phase disagreement between NR and PN results over the frequency range 
$M\omega = 0.0455$ to $M\omega = 0.1$ is greater for TaylorEt than either T1 or T4, 
but has the attractive property of decreasing monotonically as the PN order is increased. 
\end{abstract}


\maketitle

\section{Introduction}

The current interferometric gravitational-wave 
detectors~\cite{Waldman06,GEOStatus:2006,Acernese2006} 
have reached design sensitivity, and have finished taking data in the S5 science 
run. Gravitational waves 
from coalescing black-hole binaries will be among the strongest 
that one hopes to find in the detector data, and data analysts are searching for
them by performing matched filtering against template banks of theoretical 
waveforms. A broad class of theoretical templates can be produced using 
post-Newtonian (PN) approximation techniques, which are expected to be valid 
during a binary's slow inspiral. 

There exist several prescriptions to compute
gravitational-wave (GW) templates at different orders of a PN expansion. 
Each prescription, termed a PN 
approximant, provides a slightly different GW phase and hence frequency 
evolution.    
For the purpose of GW data
analysis, it is important to know which approach, at which PN order, best approximates 
the true phase evolution of GW signals from astrophysical black-hole binaries. 
If PN expansions had simple convergence properties and could be taken to 
arbitrarily high order, one would expect all approaches to converge to the same
(presumably correct)
result. However, PN calculations do not currently go beyond 3.5PN order, and the 
convergence properties are far from clear.
By 3.5PN order, we mean all the corrections up to the relative order $x^{7/2}$,
where $x = (M\, \omega_b )^{2/3}$, $M$ and $\omega_b$ are the binary's total mass
and orbital angular frequency.
 If we consider the number of GW cycles
in a given frequency window, for example, the number usually does not change
monotonically as the PN order is increased. 

One way to check the physical accuracy of the PN results is to compare
with fully general-relativistic numerical simulations over the last
several orbits of a binary's evolution. Recent breakthroughs in
Numerical Relativity (NR), reported in
Ref.~\cite{Pretorius:2005gq,Campanelli:2005dd,Baker05a}, have made it
possible to simulate many orbits before merger
\cite{Baker:2006ha,Husa2007a,Hannam:2007ik}, with the largest number
of orbits currently achieved being about 15 \cite{Boyle:2007ft}. These
simulations allow a detailed comparison with various PN approximate GW
phase evolutions.
Recent studies have shown that
standard PN approximants give good phase agreement with NR 
results up to a few orbits before merger \cite{Baker:2006ha,Hannam:2007ik,Boyle:2007ft}; 
for example, for a nonspinning equal-mass
binary, the accumulated phase disagreement between NR and PN results that use
the TaylorT1 approximant, for the seven orbits before the GW frequency reaches
$M\omega = 0.1$ (about two orbits before merger), is less than 1 radian. 

A number of PN GW template families have been proposed for use in GW searches
(see, for example, \cite{Damour00a}), and have been implemented in the 
LSC Algorithms Library (LAL) \cite{LAL} for that purpose. We focus here on
those based on the Taylor approximants. These approximants model GWs from
compact binaries inspiraling due to radiation reaction, and the phase
evolution equations are expanded in terms of $x$.
Recently one of us proposed a new class of templates \cite{Gopakumar07} in
which the phase evolution $d\phi/dt$ is now given by a PN expansion in terms of
the binding energy (see Section~\ref{sec:pn} for more details). Following the
LAL terminology, we refer to this new approximant as TaylorEt. 
The TaylorEt template has the
attractive feature that the number of accumulated GW cycles changes
monotonically as we increase the reactive PN order, which is not true for 
the other Taylor approximants. 
Further, TaylorEt templates have the same computational cost 
as, for example, the TaylorT1 and T4 approximants. 
Additionally, 
it has been recently demonstrated, while restricting radiation reaction to
dominant quadrupole contributions, that   
the TaylorEt templates
are more efficient than TaylorT1 and T4 
in capturing GWs from inspiraling compact binaries having orbital
eccentricity \cite{Tessmer07}.
Therefore, it is interesting 
to compare the GW phase evolution 
under the TaylorEt prescription
at various reactive PN orders
with the GW phase evolution in numerical simulations. 
This is what we pursue in this paper.

  In this investigation, we compare the GW phase evolution in numerical simulations
of equal-mass non-spinning black-hole binaries that last about nine orbits before 
merger with its counterparts obtainable from
TaylorT1, T4 and Et prescriptions at various PN orders.
It is important to emphasize that our NR simulations have
low initial orbital eccentricity $e \sim 0.0016$ and an accumulated numerical 
uncertainty of less than 0.25 radians in GW phase evolution. 
We observe that the accumulated GW phase difference between TaylorT1, T4 and NR
oscillates as we increase the reactive PN order from 2PN to 3.5PN in steps of 0.5PN order.
However,  differences in accumulated GW phases 
between TaylorEt and NR descriptions decrease monotonically as we 
increase the reactive phase evolution from 2PN to 3.5PN order in steps of 0.5PN order.
This implies that the least GW phase difference associated with TaylorEt occurs at 3.5PN 
order and it is $\Delta \phi = 1.18$~radians.
The observed convergence of the TaylorEt GW phase evolutions towards their 
NR counterpart and the tolerable accumulated GW phase difference for TaylorEt 
at 3.5PN order are the two main conclusions of the present paper.

In Section~\ref{sec:numerical} 
we briefly describe the numerical methods used to produce the NR waveforms. 
Section~\ref{sec:pn} details the construction of different PN prescriptions for
the GW phase evolution of the inspiral of an equal-mass binary modeled by
nonspinning point-particles. 
Section~\ref{sec:comparison}
explains how we make contact between NR and various PN descriptions for 
GW phase evolutions, and we display our results and provide 
explanations for our observations. 
Concluding remarks and future directions are given in Section~\ref{sec:summary}.

\section{Numerical methods and waveforms}
\label{sec:numerical}

The numerical simulations were performed with the BAM code
\cite{Bruegmann:2006at,Bruegmann2004}, where fourth-order accurate
derivative operators were replaced by sixth-order accurate spatial derivative 
operators in the bulk, as described in
\cite{Husa2007a}. The code started with black-hole binary puncture initial data 
\cite{Brandt97b,Bowen80} generated using a pseudo-spectral code
\cite{Ansorg:2004ds}, and evolved them with the $\chi$-variant of the
moving-puncture \cite{Campanelli2006,Baker2006} version of the BSSN
\cite{Shibata95,Baumgarte99} formulation of the 3+1 Einstein 
evolution equations. 
The gravitational waves emitted by the binary were calculated from the
Newman-Penrose scalar $\Psi_4$, using the implementation described in
 \cite{Bruegmann:2006at}. 
 
 The set of simulations that we discuss here consists of binaries that begin
 with a coordinate separation of $D = 12M$. The initial momenta are chosen
 according the prescription given in  \cite{Husa:2007ec}, which lead to 
 inspiral with a minimal eccentricity of $e < 0.0016$. For comparison, a second
 set of simulations uses initial momenta that lead to a larger eccentricity of
 $e \approx 0.008$. Simulations were performed at three resolutions. The
 results were seen to converge consistent with sixth-order accuracy and
 were subsequently Richardson extrapolated with respect to resolution. 
 Gravitational waves were extracted at five extraction radii, and the GW 
 amplitude was extrapolated with respect to extraction radius to estimate the
 GW amplitude as measured at infinity. We use the GW phase as measured at the
 largest extraction radius ($R_{ex} = 90M$). This procedure is described in 
 detail in \cite{Hannam:2007ik}, where the results of these simulations were
 first reported. 
 
 The resulting gravitational waveform has an uncertainty in the amplitude
 of less than 2\%, and an accumulated uncertainty in the phase of less than 
 0.25 radians.
In the following sections we will focus on comparing the GW phase with 
that calculated by PN methods, which we will now summarize.

\section{Various prescriptions for GW phase evolutions in PN relativity}
\label{sec:pn}

During the inspiral of a compact binary, the associated temporal GW phase evolution 
can be accurately modeled using the PN approximation to General Relativity.
In this approximation, it is 
customary to consider inspiraling astrophysical compact binaries 
undergoing adiabatic inspiral along circular orbits due 
to the emission of gravitational waves.
In recent years a number of 
computational efforts  provided
four particularly valuable PN expressions that are essential  for
GW astronomy with inspiraling non-spinning astrophysical compact binaries.
These four
quantities are the 3PN accurate dynamical (orbital) energy ${\cal E}(x)$,
expressed as a PN series in terms $x$,
the 3.5PN accurate expression for GW energy luminosity
$ {\cal L }(x) $ and the 2.5PN amplitude corrected
 expressions for $h_{+} (t)$ and $h_{\times}(t)$,
written in terms of the orbital phase $\phi$ and $x$
\cite{Damour:2001bu,Blanchet:2001ax,Blanchet:2004ek,Arun04}. 

 GW data analysis groups focusing on inspiraling compact binaries
employ these inputs to construct various types of search template.
In this paper, as mentioned earlier, we will first consider the TaylorT1
and T4 approximants.
These two template families employ the following expression 
for the restricted PN waveform
\begin{align}
h(t) \propto  x(t)^{2/3} \, \cos 2\, \phi(t)\,,
\label{EqP1}
\end{align}
where the proportionality constant
may be set to unity for non-spinning compact binaries.
At a given PN order, 
the two families provide two slightly different ways to
compute $x(t)$ and $\phi(t)$. 
To obtain the GW phase evolution $\phi(t)$ in the TaylorT1 approximant,
one  numerically solves the following two differential equations:
\begin{subequations}
\label{EqP2}
\begin{align}
\label{EqP2a}
\frac{d \phi (t)}{dt} &\equiv \omega_b (t) = \frac{c^3}{G\,M}\, x^{3/2}\,,\\
\frac{d\,x(t)}{dt} &=  -{\cal L}( x)  \left( \frac{ d {\cal E}}{d x} 
\right)^{-1}\,,
\label{EqP2b}
\end{align}
\end{subequations}
and in this section, we do not employ geometrized units.
This implies that to construct TaylorT1 3.5PN search templates, one needs to use
3.5PN accurate ${\cal L}(x)$ and 3PN accurate ${\cal E}(x)$,
respectively. The explicit expressions for these quantities,
in the case of equal mass 
compact binaries, read
\begin{subequations}
\label{EqP3}
\begin{align}
\label{EqP3a}
{\cal L}(x) &=
{\frac {2\,{c}^{5}\,}{5\,G}}\, x^5\, \biggl \{
1-{\frac {373}{84}}\,x+4\,\pi\,{x}^{3/2}-{\frac {59}{567}}\,{x}^{2}
\no
&
-{
\frac {767}{42}}\,\pi\,{x}^{5/2}
+ \biggr [
 {\frac {18608019757}{
209563200}}+{\frac {355}{64}}\,{\pi }^{2}
-{\frac {1712}{105}}\,\gamma
\no
&
-{\frac {1712}{105}}\,\ln  \left( 4\,\sqrt {x} 
\right) 
 \biggr ] {x}^{3}
+{\frac {16655}{6048}}\,\pi \,{x}^{7/2}
\biggr \}\,,
\\
{\cal E}(x) &= -\frac{M\,c^2}{8}\,x \biggl \{ 
1+ 
-{\frac {37}{48}}\,x
-{\frac {1069}{384}}\,{x}^{2}
+ \biggl [ {\frac {1427365}{331776}}
\no
&
-{\frac {205}{384}}\,{\pi}^{2}
  \biggr ]  {x}^{3}
\biggr \}\,,
\label{EqP3b}
\end{align}
\end{subequations}
where $\gamma$ is the Euler gamma.

An alternative PN approximant, TaylorT4, has recently been introduced \cite{Boyle:2007ft}.
The TaylorT4 approximant is obtained by Taylor expanding the right hand side 
of Eq.~(\ref{EqP2b}) for $dx/dt$
and truncating it at the appropriate reactive PN order.
Therefore, to construct TaylorT4 3.5PN search templates, the following
set of differential equations are numerically integrated:
\begin{subequations}
\label{EqP4}
\begin{align}
\label{EqP4a}
\frac{d \phi (t)}{dt} &\equiv \omega_b (t) = \frac{c^3}{G\,M}\, x^{3/2}\, ,\\
\frac{d\,x(t)}{dt} &= 
\frac{16\,c^3}{5\,G\,M}\, x^5\, \biggl \{
1
-{\frac {487}{168}}\,x
+4\,\pi\,{x}^{3/2}
\no
&
+{ \frac {274229}{72576}}\,{x}^{2}
-{\frac {254}{21}}\,\pi\,{x}^{5/2}
\no
&
+
\biggl [
{\frac {178384023737}{
3353011200}}-{\frac {1712}{105}}\,\gamma+{\frac {1475}{192}}\,{\pi}^{2
}
\no
&
-{\frac {856}{105}}\,\ln \left( 16\,x \right)  
\biggr ]
 {x}^{3}
+{\frac {3310}{189}}\,\pi\,{x}^{7/2}
\biggr \}
\,. 
\label{EqP4b}
\end{align}
\end{subequations}
The Taylor T1 approximant has been compared with several sets of numerical
simulations \cite{Hannam:2007ik,Boyle:2007ft} and found to agree with the
NR phase to within one radian over the GW frequency range $M\omega = 0.05$ and
$M\omega = 0.1$, when matched at $M\omega = 0.1$. (Note that the GW frequency
is related to the binary's orbital frequency by roughly $\omega = 2
\omega_b$.) The Taylor T4 approximant
has been found to agree within 0.06 radians when compared in the same way
\cite{Boyle:2007ft}. 

Let us now describe the TaylorEt approximant. The restricted PN waveform reads
\begin{align}
\label{EqP5}
h(t) & \propto {\cal  E}(t) \, \cos 2\,\phi (t) \, .
\end{align}
The time evolution for $\phi (t) $ and ${\cal  E}(t)$
are obtained by solving the following coupled differential equations.
\begin{subequations}
\label{EqP6}
\begin{align}
\label{EqP6a}
\frac{d \phi (t)}{dt} &\equiv \omega_b (t) =  
\frac{ c^3}{G\,m}\,\xi^{3/2} \biggl \{
1
+{\frac {37}{32}}\, \xi
+{\frac {12659}{2048}}\,{\xi}^{2}
\no
&
+ \biggl [
 {\frac {205}{256}}\,{\pi}^{2}+{\frac {3016715}{196608}}
\biggr ]
{\xi}^{3}
\biggr \}\,,
\\
\frac{d\,\xi (t)}{dt} &=
{\frac {16\,c^3}{5\,G\,M}}\,{\xi}^{5} 
\biggl \{
1
-{\frac {197}{336}}\,\xi
+4\,\pi\,{\xi}^{3/2}
+{\frac {374615}{72576}} \,{\xi}^{2}
\no
&
+{\frac {299}{168}}\,\pi\,{\xi}^{5/2}
+ \biggl [ 
{\frac {3155}{384 }}\,{\pi}^{2}
\no
&
-{\frac {1712}{105}}\,\ln  \left( 4\,\sqrt {\xi} \right) 
+{\frac {4324127729}{82790400}}
\no
&
-{\frac {1712}{105}}\,\gamma 
\biggr ] { \xi}^{3}
+{\frac {4155131}{96768}}\,\pi\,{\xi}^{7/2}
\biggr \}
\,.
\label{EqP6b}
\end{align}
\end{subequations}
where 
$\xi = -{2\, \cal E}/\mu\,c^2$ and $\mu$ being the usual reduced mass.
In this paper 
we keep $d \phi/dt$ 
to 3PN accuracy (and this is the highest PN order available for 
Eq.~(\ref{EqP6a}) associated with compact binaries in PN accurate circular orbits). 
The values of $\xi $ corresponding to  any initial and final GW 
frequencies can be numerically evaluated using the right-hand side of Eq.~(\ref{EqP6a}).
This is possible due to the fact that for GWs from compact binaries, having
negligible orbital eccentricities,
the frequency of the dominant harmonic is $ f_{\rm GW} \equiv \omega_b/2 $. 
 
At first sight the only difference between TaylorT1/T4 and Et is
a different choice of expansion variable. This in itself is interesting to
consider: we would like to verify that different (valid) choices of expansion
variable do not dramatically change the predictions from PN approximants.
In addition, however, the TaylorEt approximant can be easily
generalized to eccentric binaries, and indeed the lower order terms
in Eq.~(\ref{EqP6a}) are responsible for precession of an eccentric
binary. 

 It should be noted that GW phase and frequency evolutions are prescribed in certain 
parametric PN accurate ways in the TaylorEt approximant.
This is achieved in the TaylorEt approach by prescribing 
$d \phi/dt $, governing the PN accurate conservative
orbital phase evolution, in terms of ${\cal E}$ and 
then numerically imposing secular changes in ${\cal E}$,
due to the emission of GWs, with the help of far-zone GW energy flux, 
expressed in terms of ${\cal E}$.

  In the next section, we compare GW phase evolutions predicted by 
NR simulations and TaylorT1, T4 and Et approximants at various PN orders
in a given GW frequency window.

\section{Making contact between NR and PN GW phase evolutions}
\label{sec:comparison}

\begin{figure}[t]
\centering
\includegraphics[height=4cm]{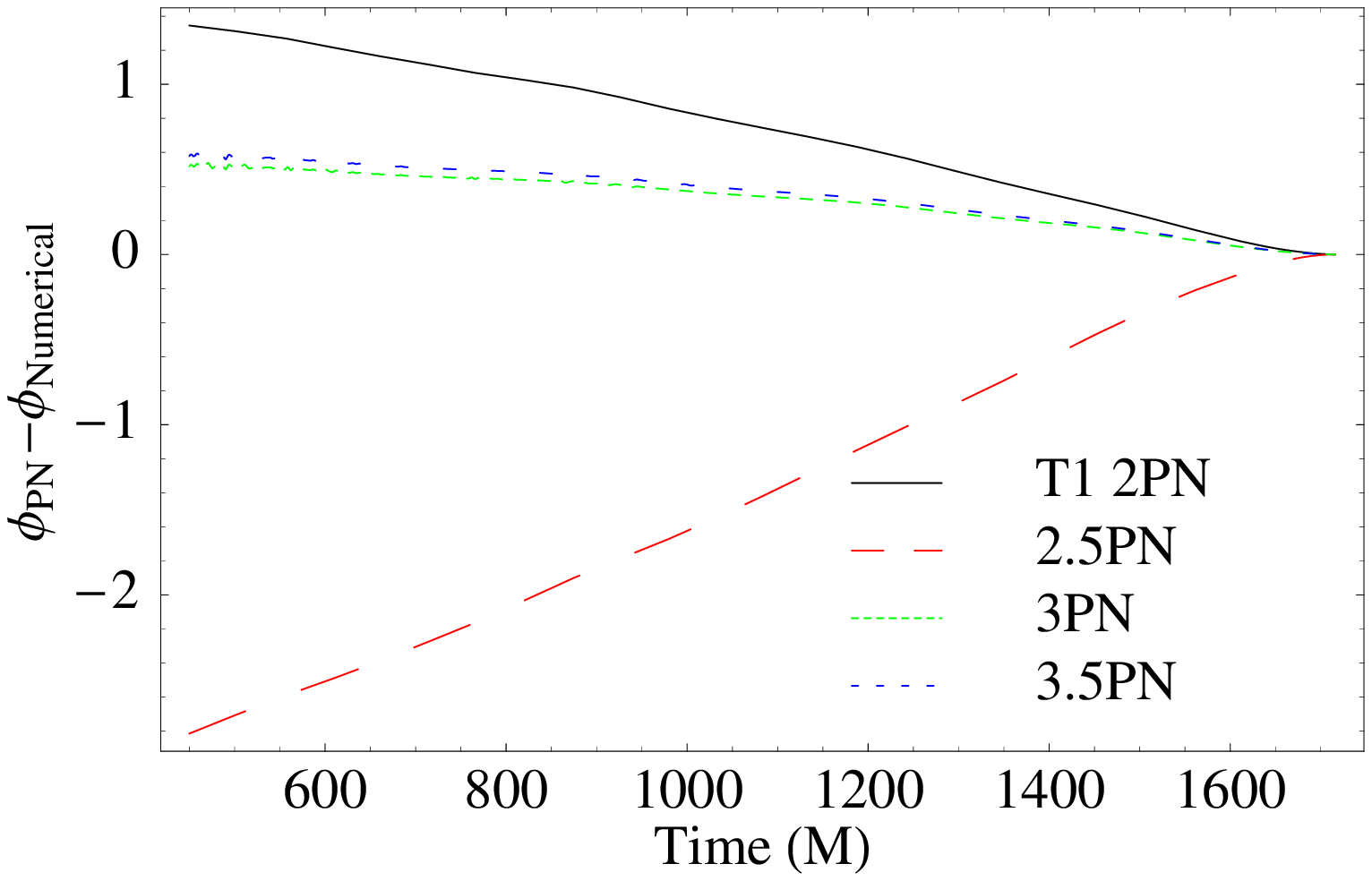}
\includegraphics[height=4cm]{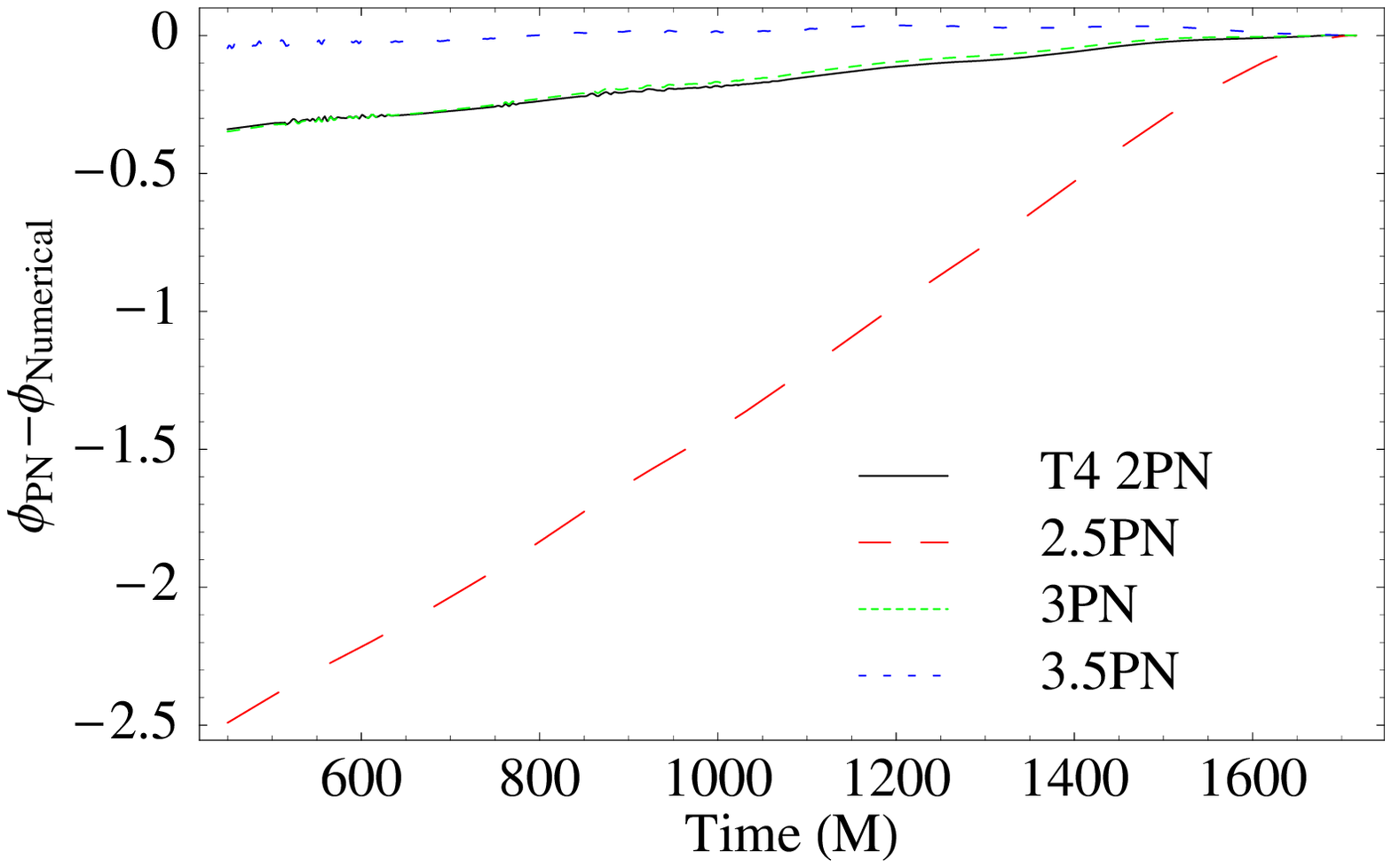}
\includegraphics[height=4cm]{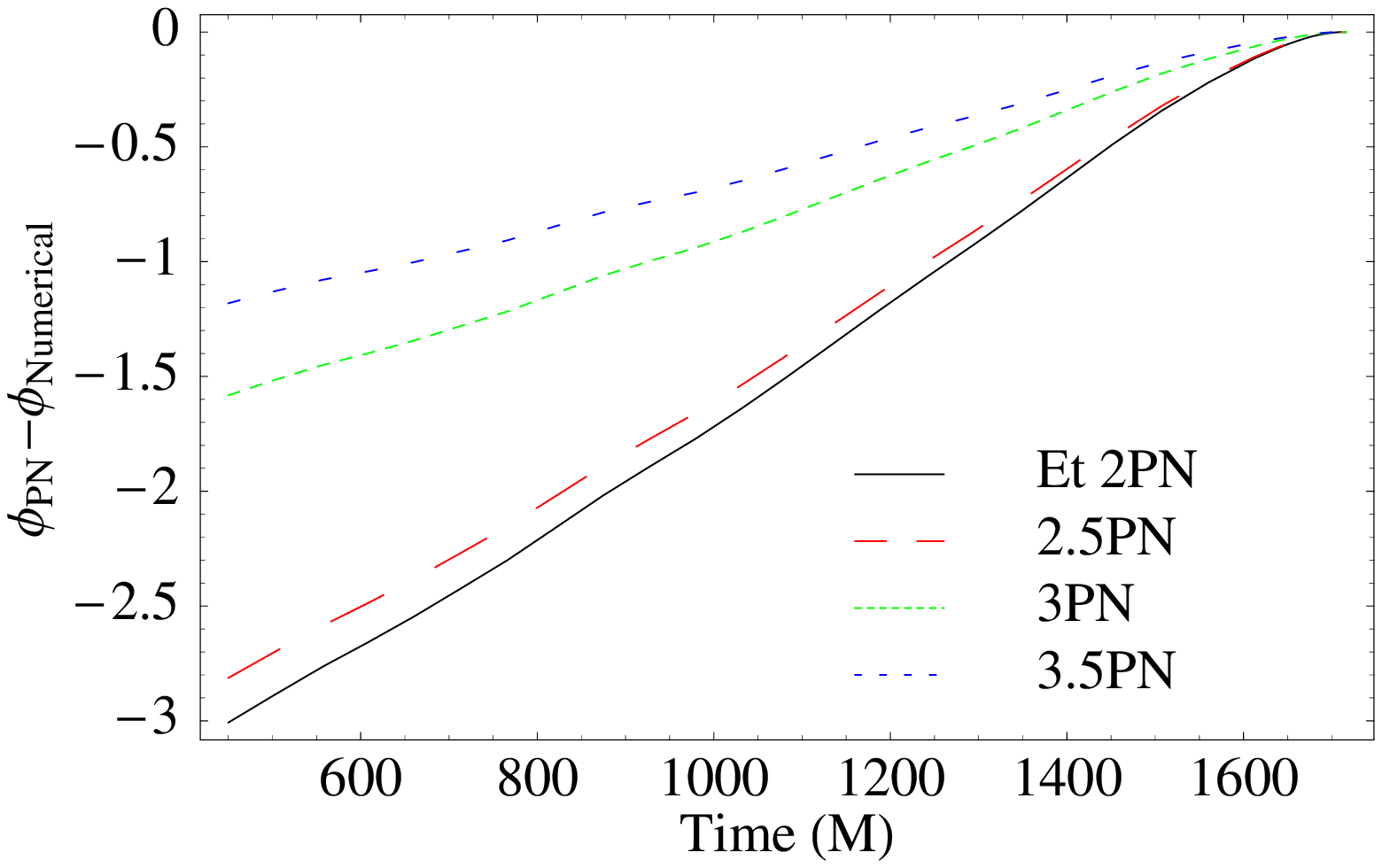}
\caption{Accumulated phase disagreement between NR and PN evolutions 
for TaylorT1, T4 and Et approximants at 2PN, 2.5PN, 3PN and 3.5PN reactive orders.}
\label{fig:Phase}
\end{figure}

 In Figure~\ref{fig:Phase}, we plot GW phase
differences, with respect to NR simulations in a given
GW frequency window,
at four post-Newtonian orders,
namely 2PN, 2.5PN, 3PN and 3.5PN, associated with the
three different PN approximants described earlier. 
The GW frequency window we employ is between $M\omega = 0.0455$ and $M\omega =
0.1$, and we line
up the NR and PN GW phases and frequencies at $M\omega = 0.1$.
The top panel refers to TaylorT1 and we observe the usual
fluctuating differences in GW
phases with respect to the PN order.
The center panel shows a comparison with the  TaylorT4 approximant. 
The good agreement at 3.5PN order that was
reported in \cite{Boyle:2007ft} is clearly visible.
Again, we observe that the PN-NR phase disagreement fluctuates with respect to
PN order. 
Further, substantially different GW phase evolutions 
in comparison with their NR counterpart
for TaylorT1 and T4 at 2.5PN
order are also observed.

In the bottom panel, we focus our attention on the TaylorEt approximant.
When we increase PN accuracy of the reactive dynamics, given by Eq.~(\ref{EqP6b}),
from 2PN to 3.5PN in steps of 0.5PN, the GW phase difference, {\it i.e.}
$\phi_{\rm PN} - \phi_{\rm NR}$, decreases in a monotonic manner.
This implies that $\phi_{\rm PN} - \phi_{\rm NR}$ 
has the
lowest value at the 3.5PN order,
and is given by $\Delta \phi = - 1.18$~radians.
This GW phase difference at 3.5PN order is definitely more than 
its counterparts arising from TaylorT1 and T4 ($\Delta \phi = 0.6$ and 0.05
respectively). 
However, it is intriguing that the TaylorEt approximant seems to display
monotonic convergence towards the NR phase. 
We have checked that monotonic convergence to NR is independent of 
the choice of our matching frequency.

  It is possible to explain, using PN based arguments, the comparatively large 
gap one finds between 2.5 and 3PN orders in the bottom panel of Fig.~\ref{fig:Phase}.
Recall that the conservative phase evolution in the TaylorEt approximant is 
3PN accurate (in other words, we always use 3PN accurate expression for
$ d\phi/dt$, given by Eq.~(\ref{EqP6a})) and this requires a complete knowledge about
the 3PN accurate conservative orbital dynamics. 
Further, it should be clear that the reactive GW phase evolution 
in TaylorEt is governed by Eq.~(\ref{EqP6b}) that prescribes $\xi (t)$.
It is not difficult to realize that to compute $d \xi/dt$ at 2PN and 2.5PN
orders, one only needs to know conservative orbital dynamics to 2PN order.
However, computations that lead to $d \xi/dt$ at 3PN and 3.5PN
orders require conservative orbital dynamics to 3PN order.
We expect that the change in the conservative dynamics from 2PN to 3PN order
while computing  $d \xi/dt$ is the dominant reason for the observed 
gap in $\phi_{\rm PN} - \phi_{\rm NR}$ between 2.5PN and 3PN orders
in the bottom panel of Fig.~\ref{fig:Phase}.
 
\begin{figure}[t]
\centering
\includegraphics[height=4cm]{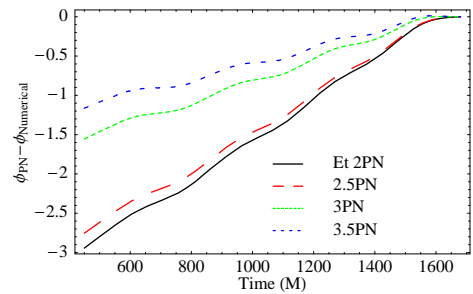}
\caption{Accumulated phase disagreement between NR results and the PN
  approximant TaylorEt, as in the lower panel of Figure~\ref{fig:Phase}, with
  the difference that we now use numerical data from a simulation with
  slightly larger eccentricity, $e \approx 0.008$.}
\label{fig:PhaseQC}
\end{figure}

 We provide Fig.~\ref{fig:PhaseQC} to argue that monotonic convergence to the
 `exact' GW  phase, exhibited by the TaylorEt templates is rather insensitive to
small numerical eccentricities present in NR evolutions.
We obtained Fig.~\ref{fig:PhaseQC} by employing NR puncture evolutions having 
an initial eccentricity $e \approx 0.008$. The oscillatory nature of 
$\phi_{\rm PN} - \phi_{\rm NR}$ clearly demonstrates that we are indeed dealing
with NR simulations that possess some tiny orbital
eccentricity.  It is also evident from Fig.~\ref{fig:PhaseQC} that 
$\phi_{\rm PN} - \phi_{\rm NR}$ at any given PN order is not that substantially 
different from what is given in the bottom panel of Fig.~\ref{fig:PhaseQC};
the accumulated phase difference was $\Delta \phi = - 1.18$~radians for the
data with negligible eccentricity, and is $\Delta \phi = -
1.16$ for the more eccentric data. 
 
 Let us explore another unique feature of TaylorEt approximant relevant for 
making contact with NR-based black-hole binary evolutions. 
A close  inspection of  Eqs.~(\ref{EqP6}) defining TaylorEt GW phase evolution
reveals that 
if one can provide bounding values for $\xi$, it is possible to 
obtain the associated $\phi(t)$ at various PN orders.

One can estimate the binding energy on one time slice of a numerical
black-hole spacetime by calculating the difference between the total energy of
the spacetime, the Arnowitt-Deser-Misner (ADM) mass, and the total mass of the two
black holes. This estimate of the binding energy was introduced in
\cite{Cook94}, and is usually performed on the initial time slice, and as such
includes the energy in the ``junk'' radiation associated
with the standard initial-data choices. However, the junk radiation quickly
radiates away, and then we may make a potentially more reliable estimate of
the binding energy by calculating \begin{equation}
E_b = E_{ADM} - M_1 - M_2 - E_{rad},
\end{equation} assuming that one has accurately measured the energy lost in
gravitational-radiation emission, $E_{rad}$. We find that we can estimate the
radiated energy with sufficient accuracy to place an uncertainty on the
binding energy of less than 2\%. 

\begin{figure}[t]
\centering
\includegraphics[height=4cm]{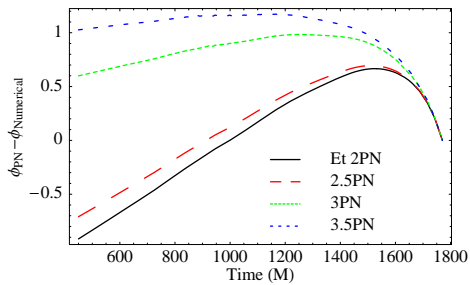}
\caption{The same comparison as in Figure~\ref{fig:PhaseQC}, but we now match
  the phase and {\it binding energy} such that $E_b = -0.01383 M$ at the matching time.}
\label{fig:PhaseQCEnergy}
\end{figure}

When we perform a matching with respect to binding energy, such that the binding energy
and phase are equal at some time (in Figure~\ref{fig:PhaseQCEnergy}, $E_b = -0.01383M$ 
for both NR and PN waveforms at the matching time), the GW frequencies of the 
NR and PN waveforms are {\it not} the same at that time. Therefore the time derivative
of the phase disagreement is non-zero, and the phase disagreement immediately grows 
linearly as we move backwards in time. 

The curves in Figure~\ref{fig:PhaseQCEnergy} possess a turning point that makes an estimate
of the accumulated phase error ambiguous. However, we should point out that this
ambiguity exists in all NR-PN phase comparisons, and this is clear in the 
figures shown in Refs.~\cite{Hannam:2007ik} and \cite{Boyle:2007ft}.  It is
always possible to perform a matching at a  
different time (and therefore a different frequency or, in this case, binding
energy), and to find a different accumulated phase disagreement,
depending on the new location of the turning point. The comparison between the
NR and PN phases is cleanest when a matching point is chosen such that there
is no turning point, as in the analysis presented earlier and demonstrated in
Figure~\ref{fig:Phase}. Despite these ambiguities, we once again see only
monotonic changes in the accumulated phase error when comparing NR results
with the TaylorEt approximant.

\section{Discussion}
\label{sec:summary}

\begin{figure}[t]
\centering
\includegraphics[height=4cm]{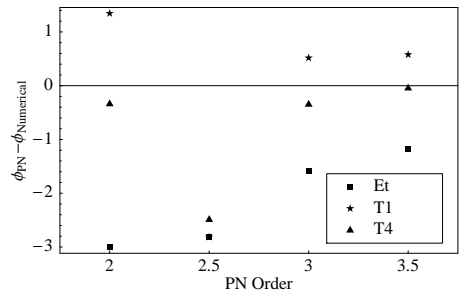}
\caption{Accumulated phase disagreement between NR and PN results, for each of the
  three approximants, TaylorT1, TaylorT4 and TaylorEt
at 2PN, 2.5PN, 3PN, 3.5PN orders. At 2.5PN the TaylorT1 and TaylorEt points
are on top of each other.
Note that the disagreement between NR and TaylorEt decreases
 monotonically as the reactive PN order is increased.}
\label{fig:PhaseSummary}
\end{figure}

 In this paper, we have compared GW phase evolutions associated with 
non-spinning equal-mass
NR based black-hole binary inspiral lasting nine orbits with three different 
prescriptions based on the TaylorT1, T4 and Et approximants.
We verified that GW phase evolution prescribed by TaylorT4 at 3.5PN order
agrees very well with its NR counterpart.
Further, $\phi_{\rm PN} - \phi_{\rm NR}$ associated with TaylorT1 and T4 
approximants fluctuates as we increase PN order from 2PN to 3.5PN in steps of 0.5PN.
However, the recently introduced TaylorEt approximant
displays an intriguing feature, namely,
$\phi_{\rm PN} - \phi_{\rm NR}$ decreases in a monotonic fashion as we increase
the PN order responsible for reactive GW phase evolution.
We also infer that $\phi_{\rm PN} - \phi_{\rm NR}$ associated with the TaylorEt 
approximant at 3.5PN order should be
 tolerable for GW data analysis purposes.
Figure~\ref{fig:PhaseSummary} summarizes our NR versus PN comparisons.

 The monotonic convergence to the fully general-relativistic GW
phase evolution, exhibited by the TaylorEt templates, along with its 
other attributes listed earlier, 
makes it an interesting candidate 
for GW templates that model inspiral, or for producing hybrid PN-NR
waveforms as in \cite{Pan:2007nw,Ajith:2007qp,Ajith:2007kx}.

The observation that TaylorEt templates should be 
more efficient in capturing GWs from compact binaries having 
small orbital eccentricities in comparison with TaylorT1 and T4,
demonstrated in Ref.~\cite{Tessmer07} while restricting the radiation reaction to
dominant contributions, 
makes it a  promising candidate to search for GWs using data from
GEO-LIGO and VIRGO.
Data analysis implications of GW polarizations $h_{\times, +} (t)$, 
evolving under the TaylorEt prescription, relevant for ground and space-based
GW interferometers are currently under investigation.  

In future work we plan to make
similar comparisons involving unequal-mass and spinning binaries.

\acknowledgments
We are  grateful to Gerhard Sch\"afer for
fruitful discussions and persistent encouragement.
This work was supported in part by
DFG grant SFB/Transregio~7 ``Gravitational Wave Astronomy''
and
the DLR (Deutsches Zentrum f\"ur Luft- und Raumfahrt) through ``LISA Germany''.
We thank the DEISA Consortium (co-funded by the EU, FP6 project
508830), for support within the DEISA Extreme Computing Initiative
(www.deisa.org).
Computations were performed at LRZ Munich and the Doppler and Kepler
clusters at the 
Theoretisch-Physikalisches Institut,
Friedrich-Schiller-Universit\"at Jena.

\bibliography{refs}

\end{document}